# Controle fino da fase das paredes de domínio em nanofitas curvas: dos bits aos nits

Gabriel Riato de Andrade Silva[1], Vagson Luiz de Carvalho Santos[2], Guilherme Henrique Rezende Bittencourt[3]

[1]UFABC    ,    [2]UFV    ,    [3]UFV

**Resumo**

Propomos um mecanismo físico de ajuste fino da fase de uma parede de domínio no entorno das direções de equilíbrio envolvendo a aplicação de um campo externo e uma corrente em sentidos contrários. Deste modo a posição da parede de domínio fica inalterada, mas sua fase pode ser ligeiramente modificada no entorno das posições de equilíbrio iniciais, aplicando o campo/corrente externos suavemente.

**1.0) Objetivo**

O equilíbrio de paredes de domínio com corrente spin polarizada e campo magnético já foi reportado como factível em [1]. Eles usaram o campo desmagnetizantes gerado por um defeito em um nanofio para fixar a posição da parede de domínio. Aqui nos propomos fazer o mesmo com a ação de um campo externo (Zeeman) e ainda ajustar a fase da P.D. . A ação de um campo externo e de uma corrente é naturalmente distinta [2], mas mostraremos analiticamente que não só um equilíbrio é possível como pequenos ajustes no entorno deste equilíbrio podem levar a aplicações interessantes.

**1.1) Motivação**

Nas últimas décadas testemunhamos um grande avanço das tecnologias de informação e comunicação, com uma demanda crescente de armazenamento de dados [3]. Do ponto de vista dos dispositivos de armazenamento, durante todo o século XX o crescimento foi exponencial, tendo nos últimos anos a tecnologia esbarrado em algumas limitações de ordem física. Aumentando a densidade de informação, os bits foram sendo armazenados em dimensões cada vez menores, tendo atingido recentemente escala de centenas para dezenas de nanômetros [4].

Nesse ponto ocorre o problema do superparamagnetismo [5], que tende a homogeneizar a magnetização de grãos menores do que isso, impossibilitando a diminuição do tamanho dos bits e dificultando então o adensamento da informação magneticamente armazenada. Uma das sugestões para contornar esse problema é a adoção de uma tecnologia baseada em outros princípios. Dispositivos de memória compostos por nanofios magnéticos foram propostos para contornar essa barreira. São chamadas racetrack memories, conceito cunhado por Stuart Parkin, pesquisador da IBM, há alguns anos [6].

Nesta tecnologia os bits de informação seriam armazenados não pela magnetização de diferentes partes do material mas pelas fronteiras entre os diferentes domínios magnéticos: as paredes de domínio [7]. O grupo de nanomagnetismo e spintrônica do DPF-UFV vem estudando a estática e dinâmica de paredes de domínio há alguns anos, tendo esclarecido alguns pontos relevantes para a construção de racetrack memories, entre eles a natureza das oscilações das paredes de domínio em nanofios curvos e as diferenças no comportamento das paredes de domínio sobre nanofios e nanofitas magnéticas curvas [8]. Paredes de domínio tendem a se mover com maior velocidade em nanofitas do que nanofios, particularmente quando se tem curvatura, nanofitas tem se mostrado uma alternativa mais viável para a propagação das paredes de domínio [9].

Correlata a dinâmica do movimento também ocorre a dinâmica da fase da parede de domínio (PD), a qual, também tem se especulado, poderia compor igualmente como meio de armazenar informação, juntamente com a ausência/presença da PD em uma região particular da nanofita. Foi mostrado que o encurvamento de uma nanofita poderia ser usado para manipular a fase de paredes de domínio que por ela transitassem [10]. Também a estática das PDs pode se mostrar interessante, tendo sido proposto um nano-oscilador baseado uma PD cuja posição se mantém fixa sobre uma nanofita devido a um degrau de anisotropia fora do plano [11].

### 1.2) Proposta

O panorama de energia em função da fase, para paredes de domínio transversais em nanofios com perfil retangular, possui duas direções de equilíbrio estável [9,10].

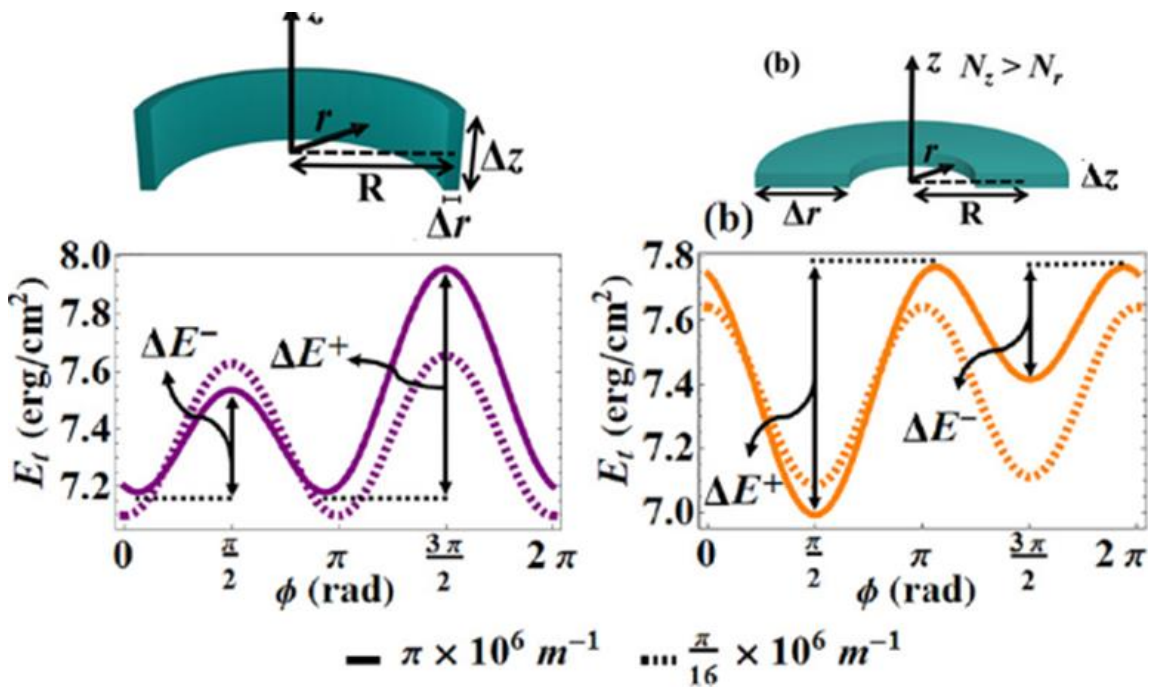

*Fig 1 - Energia x fase da parede de domínio transversal para diversos perfis de nanofita, reproduzido de [9,10].*

Um dispositivo inversor de fase, baseado na curvatura dessas nanofitas, funcionaria levando a fase da PD de um desses mínimos de energia para o outro, pela aplicação de campos e/ou correntes ao longo da direção do fio [10], como mostra a figura a seguir.

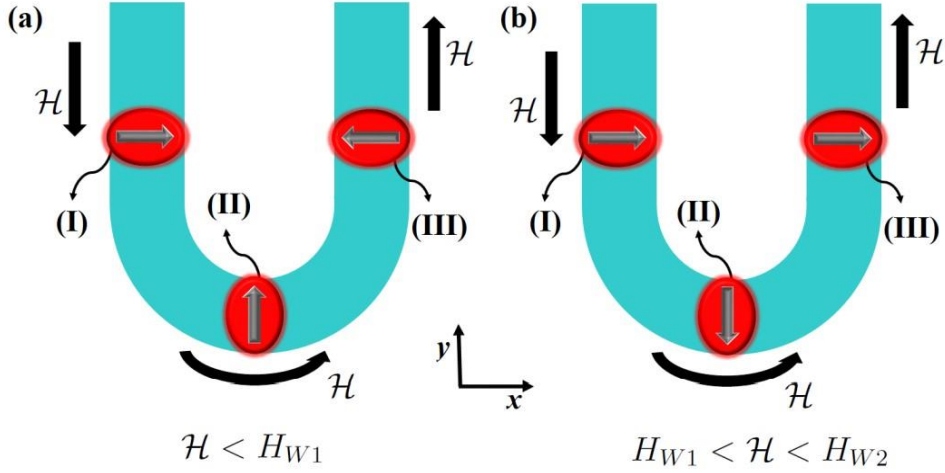

*Fig 2 - Inversão de fase em nanofitas curvas. O ajuste da composição do campo e corrente externos (𝓗) em relação aos campos de Walker ( $H_{Wi}$ ) pode definir inversões de fase. Estes campos $H_{Wi}$ correspondentes às barreiras de energia que existem entre os estados de equilíbrio. Reproduzido de [10]*

Deste modo, considerando um eventual armazenamento magnético tipo *racetrack memory* baseado em nanofitas, teríamos em cada posição de armazenamento não apenas a presença ou ausência de magnetismo, mas a ausência ou a presença em outros dois estados possíveis: magnetização na direção do primeiro ou segundo mínimo de *E(ϕ)* . O *bit* composto por 0 ou 1 cederia lugar a um *trit* composto por 0, 1 ou -1 ou 0, 1, 2 , conforme a fase da PD.

Nossa proposta aqui é criar novas posições de equilíbrio estável a partir das existentes, de modo a poder endereçar os trits de um conjunto de nanofitas, discernindo-os uns dos outros, em um buffer para *racetrack memories*. Mostraremos agora como isso pode ser obtido com o modelo teórico desenvolvido em [9,10].

### 1.3) Modelagem

A partir da equação de Landau-Lifshitz-Gilbert (1) a magnetização foi modelada conforme [9,10,12] sobre uma esfera de spin posicionada ao longo da fita, como mostrada na fig 3 (c).

$$\frac{\partial \vec{M}}{\partial t} = -\gamma \vec{M} \times \vec{H}_{eff} + \frac{\alpha}{M_S} \vec{M} \times \frac{\partial \vec{M}}{\partial t} - \frac{u}{R}\frac{\partial \vec{M}}{\partial \theta} + \frac{\beta u}{M_S R} \vec{M} \times \frac{\partial \vec{M}}{\partial \theta} \qquad (1)$$

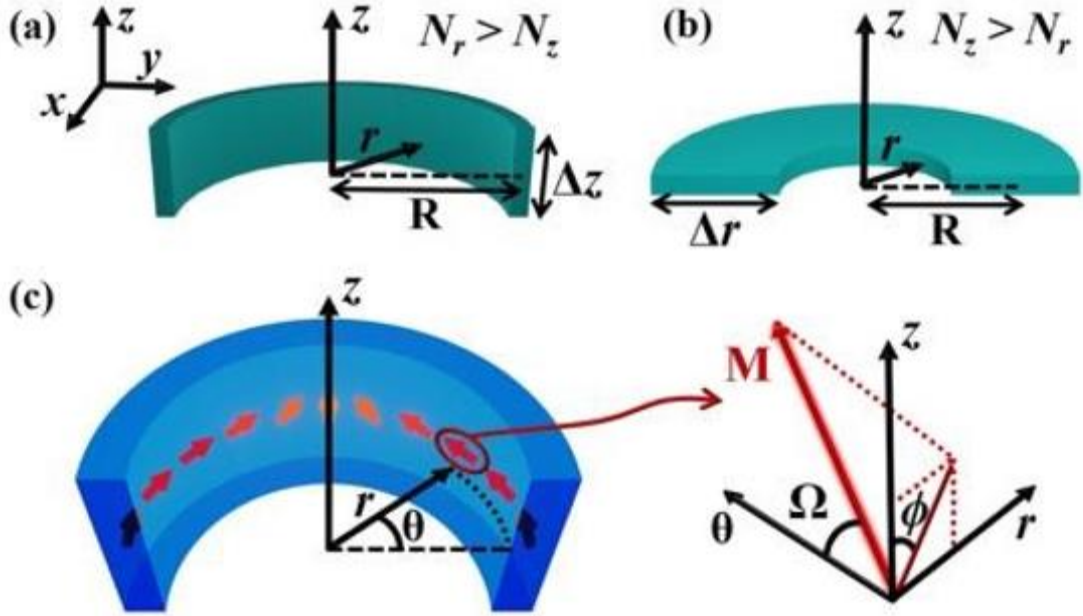

*Fig 3 - Sistema estudado: nanofitas curvas de raio médio R muito maior que dimensões transversais $\Delta r$ e $\Delta z$. Figura reproduzida de [9,10].*

Fizemos assim uma decomposição de $\vec{M}$ desta P.D. transversal como

$$\vec{m} = sen\Omega\, sen\phi\, \hat{r} + cos\Omega\, \hat{\theta} + sen\Omega\, cos\phi\, \hat{z}$$

Onde $\vec{m}$ é $\vec{M}$ normalizado pela magnetização de saturação do meio, e $\Omega = \Omega(\theta)$ depende da posição sobre a fita conforme o ansatz

$$\Omega = 2\, arctan\{\, exp[R(\theta - \theta_0)/\delta]\, \}$$

Com estas considerações na equação (1), a dinâmica do centro da parede de domínio é descrita pelo sistema de duas equações na sua fase $\phi$ e posição $q$:

$$\frac{1+\alpha^2}{\gamma} \cdot \frac{d}{dt}\begin{pmatrix}\phi \\ q/\delta\end{pmatrix} = \begin{pmatrix} 1 & \frac{\beta-\alpha}{\gamma\delta} \\ \alpha & \frac{1+\alpha\beta}{\gamma\delta} \end{pmatrix}\begin{pmatrix} H \\ u \end{pmatrix} + \begin{pmatrix} -\alpha \\ 1 \end{pmatrix} W(\phi) \qquad (1a)$$

onde

$$W(\phi) = \left[2\pi M_S(N_r - N_z) + \frac{Ak^2}{M_S}\right] \cdot sen(2\phi) - \frac{4Ak}{M_S\delta} cos\phi \qquad (1b)$$

se relaciona com os campos de Walker que definem o comportamento do sistema.

Para o armazenamento de informação nos interessa conhecer as configurações estáticas deste sistema. Um modo de obter uma configuração estática é considerar o primeiro membro nulo. Neste caso os parâmetros externos, a saber, o campo *H* e o fluxo de elétrons *u*, devem entrar em equilíbrio com os campos de Walker *W* internos do sistema. É possível mostrar que *W* ~ dE/d$\phi$, onde E é a energia da PD sem aplicação de campo externo ou corrente. Para o caso particular de campo externo e corrente nulos temos a condição *W* = 0, o que equivale a dE/d$\phi$ = 0. Estes pontos de equilíbrio devem se deslocar sobre a ação de *H* e *u* não nulos, caso em que dado um $\phi$ qualquer que se queira atingir e manter, teríamos um sistema linear em *H* e *u* direcionado por *W($\phi$)*, uma vez que esta função tomaria parte nos termos não-homogêneos do sistema.

A variável *u* se relaciona diretamente com a densidade de corrente spin-polarizada, através da taxa de polarização e constantes características como a taxa de polarização [1], por isso será eventualmente referida aqui simplesmente como *corrente*.

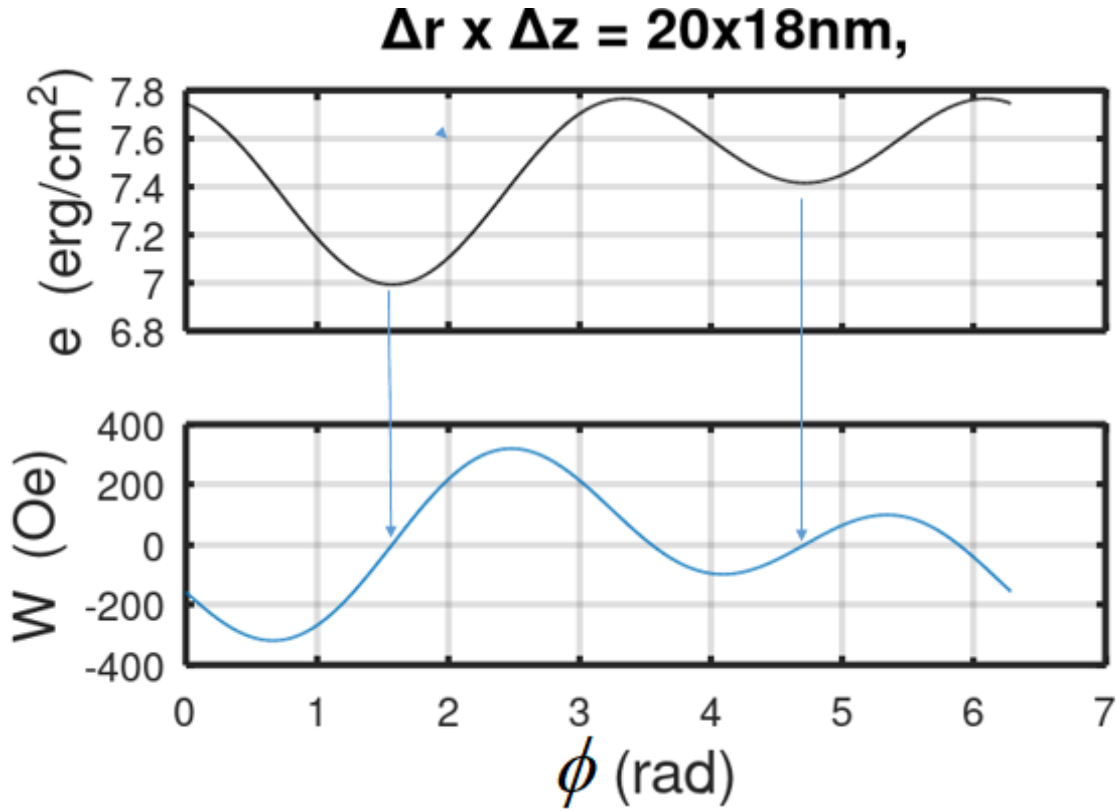

*Fig 4 – relação entre a energia por unidade de área $e(\phi)$ e o torque $W(\phi)$.*

**2. Resultados**

Para ficar claro, vamos denotar por $\phi^*$ os mínimos da energia *E($\phi$)* da PD, sem qualquer campo/corrente externo. Nesse caso a condição de equilíbrio é escrita como

$$W(\phi^*) = 0 \quad (1c)$$

Ligando *suavemente* o campo e corrente externos (e logo mais definiremos melhor o que é *suave* aqui) teremos a condição de equilíbrio descrita por $\frac{d}{dt}\begin{pmatrix}\phi \\ q/\delta\end{pmatrix} = \begin{pmatrix}0 \\ 0\end{pmatrix}$ na equação (1a):

$$\begin{pmatrix} 1 & \frac{\beta-\alpha}{\gamma\delta} \\ \alpha & \frac{1+\alpha\beta}{\gamma\delta} \end{pmatrix} \begin{pmatrix} H \\ u \end{pmatrix} = \begin{pmatrix} \alpha \\ -1 \end{pmatrix} W(\phi) \quad (2)$$

Onde $\phi$ aqui já não é exatamente $\phi^*$, mas uma posição de equilíbrio *ligeiramente* deslocada de um valor $\tilde{\phi} \equiv \phi - \phi^*$. Será pela estimativa desse valor que definiremos a transformação quase-estática que nos permitiria obter o ajuste fino da fase.

Resolvamos, por hora, resolvamos o sistema linear em no campo *H* e corrente *u* definido em (2).

$$\begin{pmatrix} H \\ u \end{pmatrix} = \begin{pmatrix} -\beta \\ \gamma\delta \end{pmatrix} W(\phi) \qquad (2b)$$

De onde, pela razão entre e primeira e segunda linhas, obtemos

$$\frac{H}{u} = -\frac{\beta}{\gamma\delta} \qquad (2c)$$

Uma vez que no membro direito de (2c) só temos constantes positivas ( o *dumping* de corrente $\beta$, o fator giromagnético $\gamma$ e a largura $\delta$ da PD) , fica provado pelo sinal negativo que, para conseguir este efeito devemos ter campo *H* e corrente *u* com sentidos opostos.

Pois bem, vamos avaliar agora quão bom pode ser nosso ajuste fino para a fase $\phi$, a partir de *H* e *u*. Para isso buscamos avaliar a incerteza da fase obtida pela aplicação de *H* e *u*. Partimos da equação (2), multiplicando a esquerda pelo vetor linha $\begin{bmatrix} \frac{1}{\alpha} & -1 \end{bmatrix}$ isolamos a função *W*:

$$W(\phi) = \frac{1}{2}\left[ \left(\frac{1}{\alpha} - \alpha\right) H + \left(\frac{\frac{\beta}{\alpha} - \alpha\beta - 2}{\gamma\delta}\right) u \right] \qquad (3)$$

Daqui vemos como a condição de equilíbrio (1c) se modificou com a aplicação do campo *H* e corrente *u* vinculados conforme (2c). Estimamos indiretamente a incerteza de $\phi$ a partir de *W*. Uma vez que *W* só depende da fase estimar sua incerteza por:

$$\sigma_W = \left|\frac{dW}{d\phi}\right| \sigma_\phi \qquad (4a)$$

Uma outra expressão para $\sigma_W$ é obtida da equação (3), por propagação dos erros[1] no campo e corrente [13]:

$$\sigma_W{}^2 = \left(\frac{\partial W}{\partial H}\sigma_H\right)^2 + \left(\frac{\partial W}{\partial u}\sigma_u\right)^2 \qquad (4b)$$

Uma outra expressão para $\sigma_W$ é obtida da equação (3), por propagação dos erros no campo e na corrente.

Note que, ao inserir (3) em (4b) obtemos um valor para $\sigma_W$ que só dependerá das constantes $\alpha, \beta, \gamma, \delta$. A partir desse valor conseguimos avaliar a incerteza na fase usando a equação (4a). Evitamos assim passar pela definição de uma inversa para função $W(\phi)$. De (4a), considerando a relação mencionada com a energia, $W \sim E'(\phi)$, resulta

$$\sigma_\phi \sim \frac{1}{|E''(\phi)|} \qquad (4c)$$

De onde extraímos uma importante interpretação. Teremos enorme dificuldade de fazer a o centro da PD apontar próximo às direções onde o gradiente de energia é mais alto ( $E''(\phi) \sim 0$ ), bem como da facilidade de direcionar a fase para regiões aonde a concavidade de $E(\phi)$ é bem definida, quaisquer que sejam os campos e correntes aplicados. Regiões com maior concavidade tem mais definidos os pontos de máximo e mínimo em $E(\phi)$, a fase é então mais direcionável no entorno desses pontos.

---

[1] A rigor esta propagação deve considerar a covariância entre o campo e a corrente, uma vez fixada a relação expressa na equação (2c), mas isto é irrelevante para a discussão qualitativa que propomos a seguir.

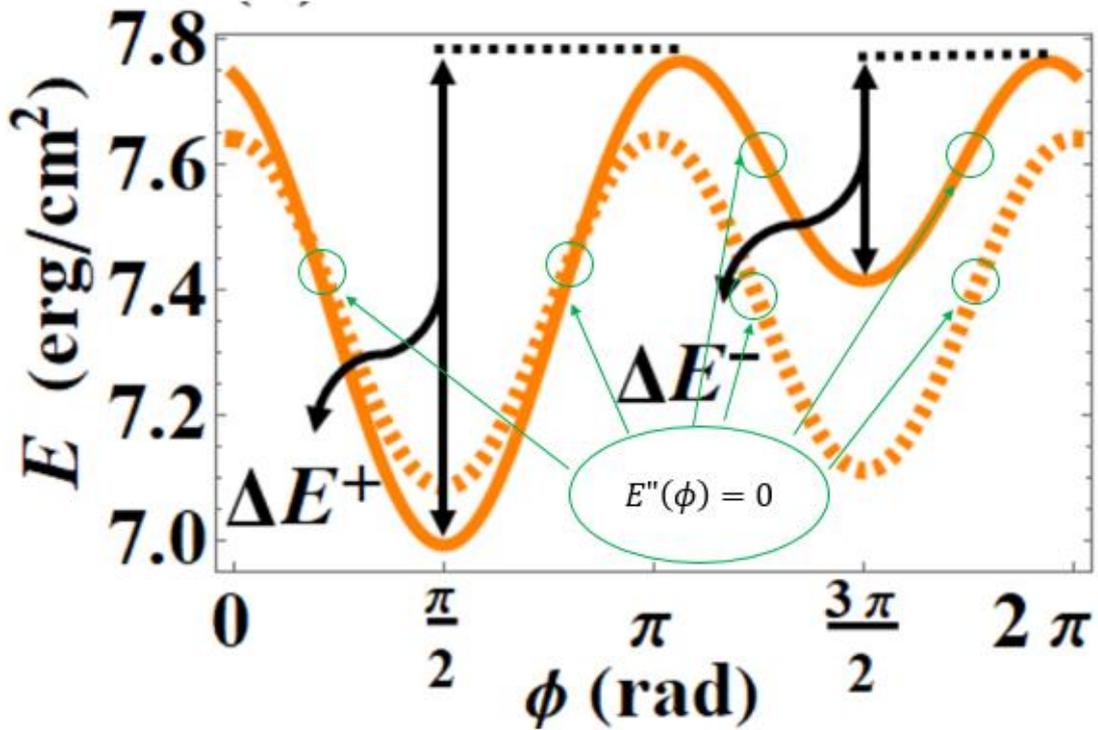

*Fig 5 - Pontos circulados são os mais instáveis da curva toda. Inexiste qualquer campo e corrente que possa fazer o centro da PD apontar nestas direções. Adaptado de [9,10].*

O cálculo da incerteza em (4b) superestima um pouco $\sigma_W^2$, pois o equilíbrio (2c) entre campo e corrente implica uma anti-correlação entre *H* e *u* que diminuiria $\sigma_W^2$[13]. Mas ela continuaria dependendo apenas das mencionadas constantes, permanecendo válido o raciocínio subsequente sobre $\sigma_\phi$. Levando em conta essa convariância entre *H* e *u*, vimos que

$$\sigma_W \geq \frac{1}{\beta}\sigma_H = \frac{1}{\gamma\delta}\sigma_u \qquad (4d)$$

Sendo que as menores incertezas seriam obtidas para valores de $\alpha$ e $\beta$ relativamente altos, como $\alpha \sim 1$ ou $\beta \sim \gamma\delta$.

Conseguimos agora delimitar mais *claramente* as condições físicas para a transformação quase-estática que nos permite controlar mais finamente a fase, no entorno dos mínimos de energia. Denotando por $\phi^{**}$ os pontos de inflexão da energia teremos

$$W'(\phi^{**}) = 0 \qquad (5)$$

Considerando o equilíbrio original (1c) podemos afirmar que o deslocamento $\tilde{\phi} \equiv \phi - \phi^*$ da fase de mínima energia causado pelo campo e corrente contrários deve ser menor do que a diferença entre um $\phi^*$ e qualquer inflexão $\phi^{**}$ adjacente a esse mesmo mínimo $\phi^*$. Em outras palavras,

$$|\tilde{\phi}| < \Delta\phi \equiv |\phi^{**} - \phi^*| \qquad (6)$$

Para evitar regiões de maior $\sigma_\phi$, consideramos razoável definir um range como

$$\phi^* - \frac{\Delta\phi}{2} < \phi < \phi^* + \frac{\Delta\phi}{2} \qquad (7)$$

para o ajuste fino da fase $\phi$.

Nesse entorno dos mínimos, pode ser razoável aproximar a energia $E(\phi)$ por série de Taylor até segunda ordem. Uma vez que $W(\phi) \sim E'(\phi)$ a função $W$ pode ser aproximada com uma expansão de primeira ordem ao redor dos pontos de equilíbrio.

$$W(\tilde{\phi}) \cong W(\phi^*) + W'(\phi^*) \cdot \tilde{\phi} \qquad (8a)$$

Sendo o primeiro termo do segundo membro nulo resulta

$$W(\phi) \cong W'^* \cdot (\phi - \phi^*) \qquad (8b)$$

O que, substituindo em (3), nos dá o deslocamento da fase em termos do campo/corrente:

$$\phi \cong \phi^* + \frac{1}{2W'^*}\left[\left(\frac{1}{\alpha} - \alpha\right)H + \left(\frac{\frac{\beta}{\alpha} - \alpha\beta - 2}{\gamma\delta}\right)u\right] \qquad (9)$$

Aqui $W'^*$ define quão profundo é o poço de potencial no qual estamos operando. Essa constante depende apenas dos parâmetros geométricos, como curvatura e secção transversal. Sua definição fica clara se reescrevermos (1b) na forma

$$W(\phi) = \widetilde{H_d} \cdot sen(2\phi) - H_{t1} \cdot cos\phi \qquad (10)$$

usando os valores máximos dos campos dipolar e de troca dados por

$$H_d \equiv 2\pi M_S(N_r - N_z) \qquad (10a)$$

$$H_{t1} \equiv \frac{4Ak}{M_S\delta} \qquad (10b)$$

$$H_{t2} \equiv \frac{Ak^2}{M_S} \qquad (10c)$$

$$\widetilde{H_d} \equiv H_d + H_{t2} \qquad (10d)$$

Deste modo,

$$W'^* = \left(1 + \frac{1}{2}\frac{H_{t1}}{\widetilde{H_d}}\right)\left(2\widetilde{H_d} - H_{t1}\right) \qquad (11)$$

Conseguimos agora definir as condições sobre o campo/corrente. Escrevendo apenas a primeira equação do sistema (1a):

$$\frac{1+\alpha^2}{\gamma} \cdot \frac{d\phi}{dt} = \mathcal{H}(H, u) - \alpha W(\phi) \qquad (12)$$

com

$$\mathcal{H}(H, u) \equiv H + \frac{\beta - \alpha}{\gamma\delta}u \qquad (12a)$$

Naturalmente queremos, uma vez atingida a nova fase de equilíbrio, que $\frac{d\phi}{dt}$ se anule. Igualando então (12) a zero e substituindo (8b) temos

$$\mathcal{H}(H, u) \cong \alpha W'^* \cdot (\phi - \phi^*) \qquad (12b)$$

Levando em conta as restrições impostas por (6) teremos

$$\mathcal{H}(H,u) < \alpha W'^*.\Delta\phi \quad (12c)$$

Lembrando que $\Delta\phi = |\phi^{**} - \phi^*|$, onde $\phi^{**}$ correspondente ao ponto de inflexão mais próximo da fase de mínima energia $\phi^*$. Pela fig 5 teríamos $\Delta\phi \sim \pi/4$ para o primeiro mínimo.

Adotando o range (7) em (12c) e considerando a definição (12a) garantiremos o ajuste fino da fase no intervalo de campo/corrente:

$$-\alpha W'^*.\frac{\Delta\phi}{2} < H + \frac{\beta-\alpha}{\gamma\delta}u < \alpha W'^*.\frac{\Delta\phi}{2} \quad (12c)$$

Caso este $\mathcal{H}(H,u) = H + \frac{\beta-\alpha}{\gamma\delta}u = \frac{\alpha}{\beta}H = \frac{\alpha}{\gamma\delta}u$ [ por (2c) ] seja atingido devemos assegurar que isso ocorra em um tempo $\Delta t$ grande suficiente, de forma quase estática.

Um tempo característico de precessão pode ser obtido pela seguinte integral de (12):

$$T[\mathcal{H}(H,u)] \equiv \int dt = \frac{1+\alpha^2}{\gamma}.\int_0^{2\pi} \frac{d\phi}{\mathcal{H}(H,u)-\alpha W(\phi)} \quad (13)$$

Uma vez que a constante $\gamma$ assume valores relativamente altos ( $\gamma \cong 1,7.10^7 Hz/gauss$ para o permalloy), ainda que o tempo $\Delta t$ da nossa mudança quase-estática de $\phi^*$ a $\phi$ seja duas ou três ordens de grandeza maior, ainda teremos um fenômeno rápido para os padrões da escala macroscópica. Como pensamos ser razoável considerar que o sistema atinja no máximo um $\phi \sim \Delta\phi/2$, num regime precessional o tempo para isso seria apenas uma fração de $T$. O gráfico a seguir mostra o comportamento de $T$ em função do campo externo $\mathcal{H}(H,u)$ conforme a equação (13):

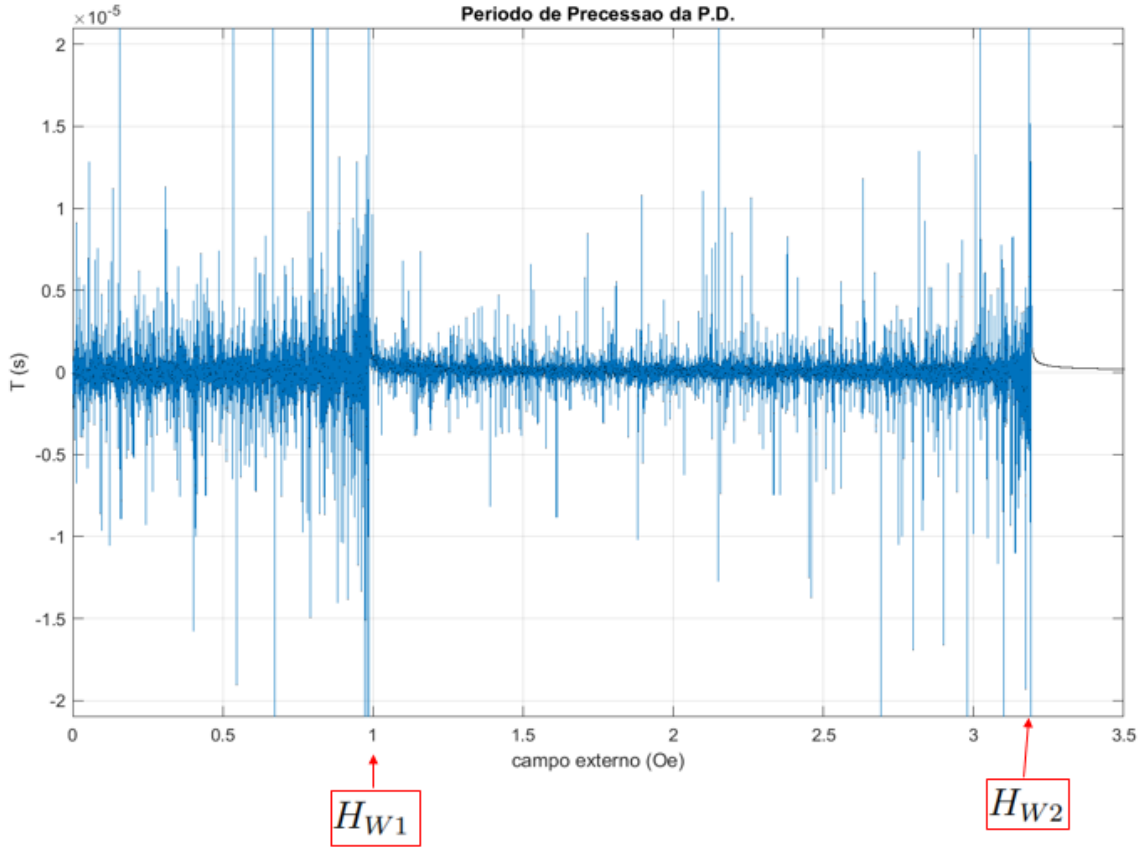

*Fig 6 - Período definido pela equação (13) em função de H para corrente nula.*

Note como inicialmente não parecemos ter um regime precessional definido, variando a integral selvagemente na região $0 < \mathcal{H} < 3,2\ Oe$, assumindo inclusive valores negativos. É fácil interpretar este gráfico tendo em vista os resultados anteriores de Bittencourt reapresentados na figura a seguir:

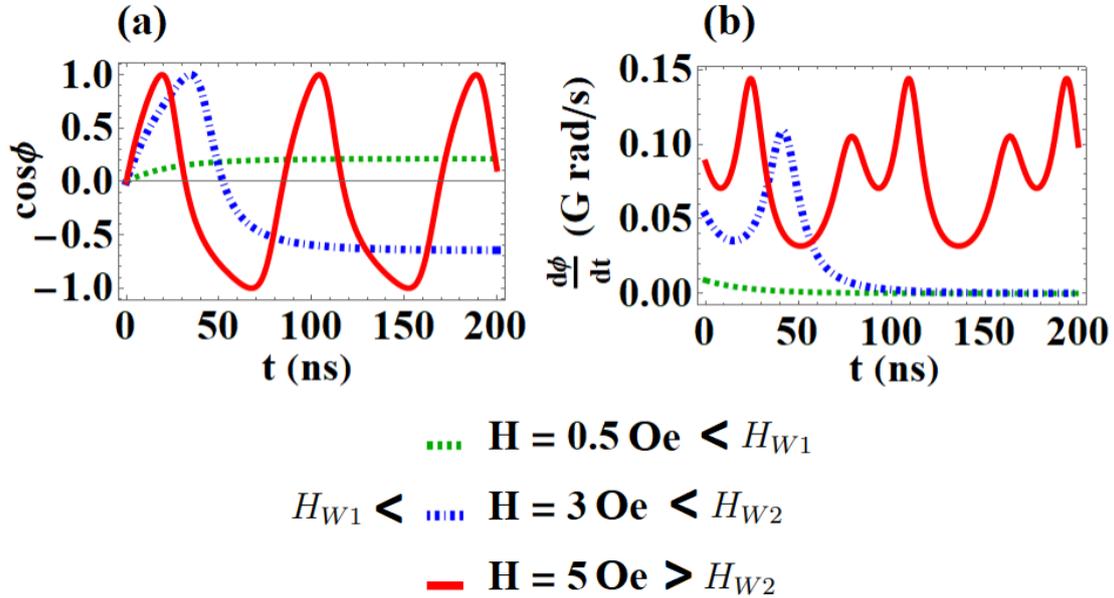

*Fig 7 - Integração das equações de movimento da parede transversal em uma nanofita. Indicado o primeiro campo de Walker. Adaptado de [9,10].*

Aqui vemos em maior detalhe o funcionamento do inversor de fase apresentado, conforme o campo externo varia em relação aos campos de Walker $H_{w1}$ e $H_{w2}$. Valores de H inferiores a qualquer um destes limites não definem precessão. Mas valores entre $H_{w1}$ e $H_{w2}$ definem a inversão mostrada pela curva azul pontilhada. Já a curva verde mostra a estabilidade da fase quando $|H|<H_{w1}$. É possivelmente nesta última janela de valores de campo que as nossas transformações quase-estáticas tem a chance de acontecer.

Essas mesmas transformações podem ocorrer sob condições de anisotropia ou campo externo não tangencias ao nanofio.

Verificamos que, introduzindo nas equações de Landau-Lifshitz-Gilbert (1), anisotropias de eixo fácil nas direções r ou z a equação (1a) mantém a forma, com pequenas correções na função $W$, conforme a constante de anisotropia $K$ aplicada. Por isso vamos considerar $W$ escrita na forma da equação (10):

$$W(\phi) = \widetilde{H_d}.sen(2\phi) - H_{t1}.cos\phi$$

Com anisotropia z temos uma diminuição no campo $\widetilde{H_d} \rightarrow \widetilde{H_d}^z = \widetilde{H_d} - \frac{1}{2}\frac{\gamma K}{M_S}$.

Já com anisotropia em r temos um aumento de $\widetilde{H_d} \rightarrow \widetilde{H_d}^r = \widetilde{H_d} + \frac{1}{2}\frac{\gamma K}{M_S}$.

Note que o aumento ou diminuição do primeiro coeficiente ocorrre na mesma proporção, dependendo apenas da direção da anisotropia considerada.

Bittencourt 2023 [14] considera nanofitas curvas sob a ação de um campo tangencial cartesiano, como mostrado na figura a seguir, constantando oscilações análogas às de um pêndulo simples.

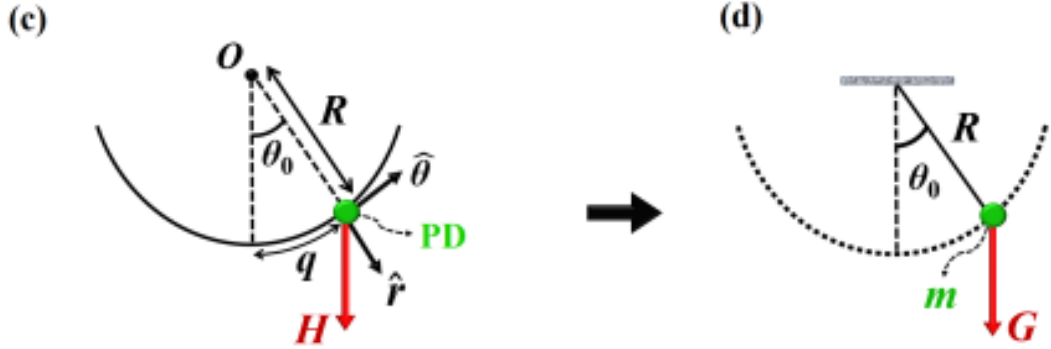

*Fig 8 - Sistema com parede de domínio transversal apresentando comportamento oscilatório. Retirado de [14].*

Repare que, neste caso, o campo externo aplicado é cartesiano e ele procede pela linearização das equações no regime de pequenas oscilações. Isto não seria necessário para constatar também neste caso a dirigibilidade quase-estática da fase que propomos aqui. Para resumir, se escrevermos a integração que nos leva a energia do campo Zeeman temos

$$E_Z = \frac{1}{2}\int \vec{M}\cdot\vec{H}\,SRd\theta,$$ com $\vec{H} = H(cos\theta, -sen\theta, 0)$ em coordenadas cilíndricas.

Pois o elemento de volume da nanofita é $SRd\theta$, com $S$ a área da secção transversal. Decompondo a magnetização desta PD transversal e a descrevendo pelo mesmo ansatz $\Omega(\theta)$ chegamos a

$$E_Z = \frac{1}{2}SRMH\int (sen\Omega.cos\theta.sen\phi - cos\Omega.sen\theta)\,d\theta$$

Portando essa energia tem a forma geral:

$$E_Z(\phi) = A.sen\phi + B \qquad (14)$$

Com $A$ e $B$ constantes. Somando-a na energia total:

$$E = E_d + E_t + E_Z \Rightarrow E(\phi) = E_0(\phi) + E_Z(\phi) \Rightarrow E'(\phi) = E'_0(\phi) + E'_Z(\phi)$$

Onde $E_0(\phi) = E_d(\phi) + E_t(\phi)$ é a energia da PD sem campo/corrente externo. Como a derivada dessa energia em relação fase é ~$W$ e derivando (14) temos $E'_Z = A.cos\phi$, podemos dizer que existe então uma outra $W$ correspondente a este caso.

$$E' = E'_0 + E'_Z \Rightarrow W = W_0 + W_Z \text{ com } W_Z \equiv H_Z.cos\phi\,, H_Z \sim H$$

E, de modo análogo a anisotropia, a nova $W$ só difere pela modificação de um coeficiente. Ou seja, com o campo transversal meramente trocamos $H_{t1} \rightarrow \widetilde{H_{t1}} = H_{t1} - H_Z(H)$, obtendo

$$W(\phi) = \widetilde{H_d}.sen(2\phi) - \widetilde{H_{t1}}.cos\phi \qquad (15)$$

Que é totalmente análoga a (1b) e (10).

### 3. Discussão dos resultados

Considerando um $\Delta\phi \sim \pi/4$ na expressão (12c) estimamos valores máximos de $|\mathcal{H}(H,u)|$ compatíveis com o intervalo $0 < \mathcal{H} < 3,2\ Oe$ correspondente ao regime não precessional indicado no gráfico $T$ x $\mathcal{H}$. Expressões semelhantes a (12c) poderiam ser igualmente obtidas por considerações termodinâmicas.

A eventual diminuição do segundo coeficiente de *W*, introduzida pela ação de um campo transversal cartesiano, pode levar a um alargamento do primeiro poço de potencial da nanofita, eventualmente cedendo mais espaço para as transformações quase-estáticas que propomos visando o controle fino da fase no entorno dos mínimos de energia.

### 4. Conclusões

Com base nessas transformações quase-estáticas podemos propor uma múltipla codificação para um *buffer* de memória *racetrack* constituída por nanofitas curvas. As fases quase-estaticamente deslocadas atribuiriam endereçamento e diferenciam a informação de cada nanofita dentro do conjunto.

A quantidade de nanofitas neste *buffer* é tão maior quanto menores forem as incertezas $\sigma_\phi$ conseguidas para o ajuste de fase em cada nanofita. Como mencionado, tais incertezas não dependeriam apenas de $\sigma_H$ e $\sigma_u$ mas também de uma escolha cuidadosa do material ( constantes $\alpha$ e $\beta$ ).

A figura a seguir esquematiza um *buffer* constituido por um feixe de 3 nanofitas. As fases de repouso de cada uma dessas nanofitas estão deslocadas por uma tolerância discriminatória $Z_\phi \sim \sigma_\phi$ , que poderia ser de duas, três, quatro ou até mais incertezas, dependendo quão pequena for $\sigma_\phi$ em relação a $\Delta\phi$ e quanto gostaríamos de minimizar eventuais erros de leitura.

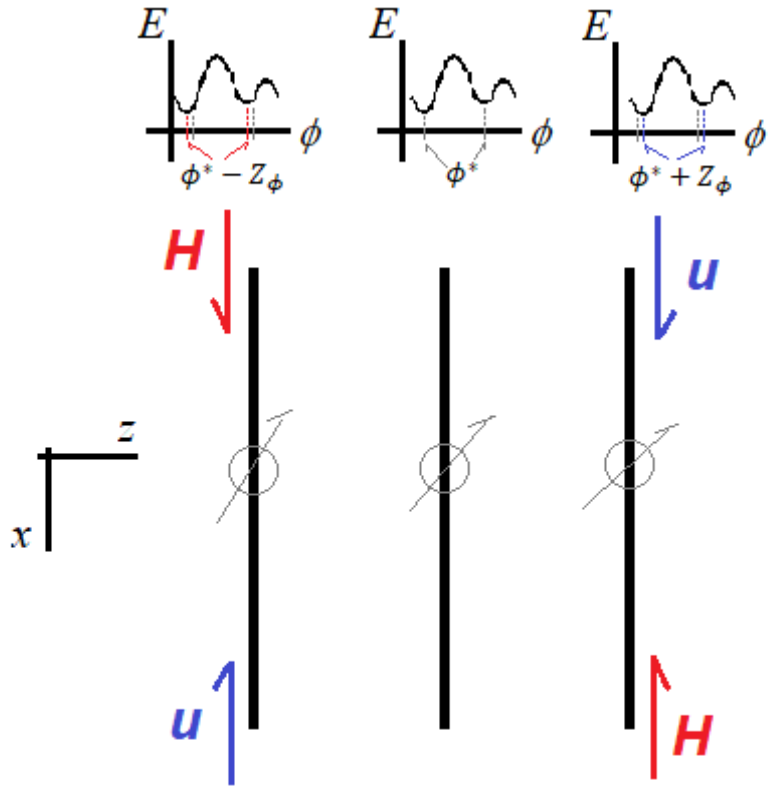

*Fig 9 – buffer de memória a nanofitas com endereçamento por deslocamento de fase quase-estático. Posiçoes de equilíbrio a direita deslocadas de $Z_\phi$ em relação ao fio do meio. Fio da esquerda tem fases deslocadas em sentido oposto ($-Z_\phi$).*

Vemos então três nanofitas com diferentes mínimos, no centro, a esquerda e a direita. Possibilitando sua diferenciação no momento da leitura através do uso conjugado de campo e corrente, conforme deduzido na equação (2c). Cada nanofita tem três estados diferentes, permitindo as transformações quase-estáticas criar ainda mais estados em cada uma delas, eventualmente ladeando a nanofita central com mais pares de outras, em fases deslocadas na proporção $\pm 2\sigma_\phi$, $\pm 3\sigma_\phi$, $\pm 4\sigma_\phi$, etc. Deste modo, do *trit* que tínhamos para cada PD sobre a nanofita (estado nulo, primeiro e segundo mínimos), podemos a princípio modificar os estados não nulos para representar outros *n* estados, a depender da resolução $Z_\phi$ obedecer o critério

$$n.\sigma_\phi < \Delta\phi/2$$

para garantirmos que os novos estados permaneçam no entorno do mesmo mínimo $\phi^*$ original.

Considerando um $\Delta\phi \sim \pi/4$ e adotando uma tolerância discriminatória $Z_\phi = 3\sigma_\phi$, com os valores típicos de campo e corrente mencionados em [1] chegamos[2] às incertezas $\sigma_\phi = \pi/24$, $\sigma_H = 0{,}82\ Oe$ e $\sigma_u = 15\ cm/s$, o que implicaria em incertezas relativas menores do que 80% para o campo e 1% para a corrente, parecendo portanto ser mais difícil ajustar a corrente certa para equilibrar a parede de domínio.

---

[2] Aqui consideramos a covariância entre *H* e *u* calculada a partir da equação (2c).

**5. Perspectivas**

A inspiração deste trabalho veio da teoria de sistemas e engenharia de controle [15-17]. Tópicos mais usualmente abordados nos cursos de computação e automação, sendo muito mais do *metier* da engenharia do que da física. Por isto mesmo agregam novos *insigths*, principalmente aos mais interessados em aplicações tecnológicas. São frameworks matemáticos para dispor as leis da natureza ao uso humano.

Controlabilidade é fundamental para qualquer tecnologia. Alguns físicos propuseram sua própria versão de uma matriz de controlabilidade para um dispositivo spintrônico [18]. A teoria do controle linear já tem algumas décadas [16] e basicamente se utiliza de matrizes para entender e classificar sistemas do ponto de vista do controle. Daí nossa inspiração para escrever as equações da parede de domínio transversal na forma (1a). Essa forma, contudo, ainda não é linearizada. A descrição de um dispositivo pela teoria de sistemas linear é feita por 4 matrizes, as quais vamos chamar aqui **A**, **B**, **C** e **D**. Tal sistema, ao receber um vetor de input x(t) devolveria um output y(t) conforme

$$\mathbf{x}'(t) = \mathbf{A}.\mathbf{x}(t) + \mathbf{B}.\mathbf{u}(t) \quad (16a)$$

$$\mathbf{y}(t) = \mathbf{C}.\mathbf{x}(t) + \mathbf{D}.\mathbf{u}(t) \quad (16b)$$

A última equação da a resposta instantânea, enquanto a primeira a resposta dinâmica (memória). O vetor u(t) contém as forças externas que atuam sobre o sistema, tentando direciona-lo e controlá-lo. Por esse motivo B e D representam os *atuadores* do nosso dispositivo, enquanto A e C representam os *sensores*. No nosso caso **u** seria $\begin{pmatrix}H\\u\end{pmatrix}$ e, comparando (16a) com (1a) vemos que **B** ~ $\begin{pmatrix}1 & \frac{\beta-\alpha}{\gamma\delta}\\\alpha & \frac{1+\alpha\beta}{\gamma\delta}\end{pmatrix}$ e **x** ~ $\begin{pmatrix}\phi\\q\end{pmatrix}$. Pensando no controle dessas variáveis dinâmicas a comparação faz sentido mas também poderíamos definir **x** ~ $\frac{d}{dt}\begin{pmatrix}\phi\\q\end{pmatrix}$, pois caso nosso interesse seja um dispositivo nano-oscilador como os propostos em [2,11], faríamos então a comparação de (1a) com (16b), obtendo **D** ~ $\begin{pmatrix}1 & \frac{\beta-\alpha}{\gamma\delta}\\\alpha & \frac{1+\alpha\beta}{\gamma\delta}\end{pmatrix}$.

As matrizes-sensoras **A** e **C** seriam definidas pela linearização de $W(\phi)$ como escrevemos nas equações (8a) e (8b), elas estariam então conectadas a profundidade do poço de potencial definido por $W'^*$.

Pensando em diferentes sistemas como o inversor da fig 2 ou o oscilador da fig 8, podemos, com esta análise linear, comparar seu comportamento dinâmico e até propor novas possibilidades escrevendo seus sistemas de equações linearizadas:

$$\frac{d}{dt}\begin{pmatrix}\tilde{\phi}\\q/\delta\end{pmatrix} = \frac{\gamma W'^*}{1+\alpha^2}\begin{pmatrix}\alpha & 0\\1 & 0\end{pmatrix}\begin{pmatrix}\tilde{\phi}\\q/\delta\end{pmatrix} + \frac{\gamma}{1+\alpha^2}\begin{pmatrix}1 & \frac{\beta-\alpha}{\gamma\delta}\\\alpha & \frac{1+\alpha\beta}{\gamma\delta}\end{pmatrix}\begin{pmatrix}H\\u\end{pmatrix} \quad (17a)$$

$$\frac{d}{dt}\begin{pmatrix}\phi\\q\end{pmatrix} = \frac{\gamma}{2M_S(1+\alpha^2)}\left\{\begin{pmatrix}-\frac{\alpha}{\lambda}\varkappa_1 & -\varkappa_2\\ \varkappa_1 & -\alpha\lambda\varkappa_2\end{pmatrix}\begin{pmatrix}\phi\\q\end{pmatrix} + \frac{\pi}{2}\begin{pmatrix}\frac{\alpha}{\lambda} & 0\\ -1 & 0\end{pmatrix}\begin{pmatrix}\varkappa_1\\ \varkappa_2\end{pmatrix}\right\} \quad (17b)$$

$$\frac{d}{dt}\begin{pmatrix}\phi\\q\end{pmatrix} = \frac{\gamma}{2M_S(1+\alpha^2)}\left\{\begin{pmatrix}-\frac{\alpha}{\lambda}\varkappa_1 & -\varkappa_2\\ \varkappa_1 & -\alpha\lambda\varkappa_2\end{pmatrix}\begin{pmatrix}\phi\\q\end{pmatrix} + \frac{\pi}{2}\begin{pmatrix}\frac{\alpha}{\lambda} & 0\\ -1 & -\frac{4M_S}{\pi\gamma}(1+\alpha^2)\frac{\lambda}{\delta}\end{pmatrix}\begin{pmatrix}\varkappa_1\\ u\end{pmatrix}\right\} \quad (17c)$$

Todas estas equações estão na forma (16a). Em (17a) e (17b) estão o inversor da fig 2 e o oscilador da fig 8. Note a coluna de zeros na matriz sensora **A** de (17a) e na matriz atuadora **B** de (17b). Isso certamente tem um impacto sobre a controlabilidade de cada um destes sistemas. Controlar a fase e a posição no inversor da fig 2 é como domar um touro, que pula e rodopia mais rápido do que podemos eventualmente sequer perceber. Já o oscilador apresentado na fig 8 é como um carro sem volante, (17b) nos mostra que tem tudo o necessário para andar mas falta um atuador essencial para garantir o controle da direção. A partir disto adicionamos um torque de corrente spin-polarizada tangencial ao fio sobre este último sistema, como representado na fig 10.

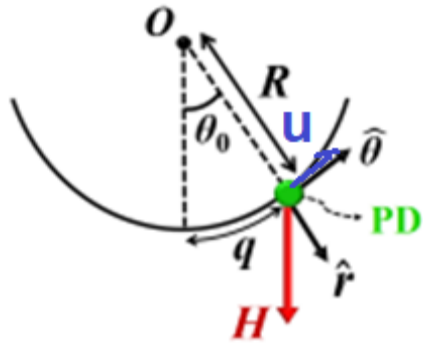

*Fig 10 – Sistema representado na fig 8, acrescido de corrente spin polarizada tangencial u. Adaptado de [14].*

É este sistema que representamos na equação (17c). Supomos que (17a), (17b) e (17c) estejam em ordem crescente de controlabilidade. Na engenharia de controle linear uma das ferramentas para indicar o grau de controlabilidade é a matriz gramiana W definida pela equação (18).

$$\mathbf{W}(t) \equiv \int_0^t \exp(\mathbf{A}\tau)\,\mathbf{B}\mathbf{B}'\exp(\mathbf{A}'\tau)\,d\tau \quad (18)$$

Onde $\mathbf{A}'$ seria a transposta de $\mathbf{A}$ e $\mathbf{B}'$ a de $\mathbf{B}$. Aqui a medida de controlabilidade seria $|\det \mathbf{W}|(t)$. Diz-se que um sistema com det**W** nulo não é controlável. Como neste caso teremos um det**W** com dimensão de frequência ao quadrado, poderíamos ainda associar det**W** positivos ou negativos a controles com oscilação ou dissipação.

Mas voltando a equação (16a), nosso vetor motriz **u**(t) foi definido como $\begin{pmatrix}H\\u\end{pmatrix}$. Note que não consideramos até aqui direcionar a parede de domínio com campos ou correntes variando no tempo. Há uma parte da engenharia de controle chamada *Teoria da Função Resposta* que se utiliza de transformadas de laplace para fazer outros tipos de análise. Por hora, foquemos no eventual significado físico de um **u**(t) = $\begin{pmatrix}H(t)\\u(t)\end{pmatrix}$ com funções oscilatórias: entramos no domínio da óptica direcionando o magnetismo [19].

Campos e correntes estáticos na fig 9, ainda que injetados balanceadamente (conforme a equação (2c)), eventualmente deslocariam do mesmo modo a fase de todas paredes de domínio que na mesma nanofita se encontrassem. Contudo, havendo uma maneira de criar algum tipo de onda estacionária sobre uma única P.D., levaríamos o controle da fases nessas paredes de domínio a outro nível.

Adentramos aqui uma especulação sobre possíveis metamateriais magnéticos com propriedades compatíveis a esta finalidade. Ondas eletromagnéticas longitudinais foram reportadas como possíveis no metamaterial elaborado por Denis Sakho e colaboradores em 2021[20]. A fig 11 esquematiza uma eventual combinação de ondas longitudinais capaz de provocar uma onda estacionária sobre uma única PD-alvo.

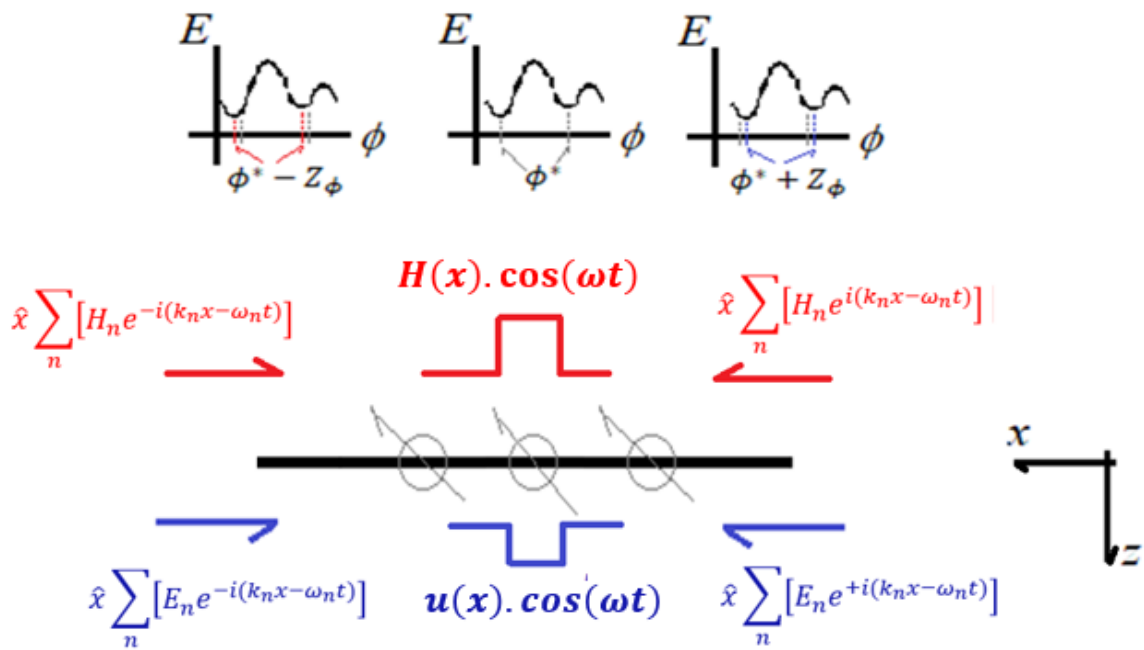

Fig 11 – ao inserir injetar eletromagnéticas longitudinais contrapropagantes na nanofita pode-se eventualmente obter uma onda estacionária que manipule a fase de uma única parede de domínio, em meio a várias outras.

A referência [20] descreve essas ondas em pulsos ultra curtos, o que usualmente se traduz em alta frequencia. No nosso caso temos como requisito baixa frequencia destas ondas, uma vez que a manipulação quase-estática que descrevemos para a fase demanda isso. Para tanto a velocidade de propagação das ondas deve ser diminuida. Isso também é factível, uma vez que já foi reportado há mais de 10 anos em materiais com alto índice de refração [21]. Talvez a união destas características com ferromagnetismo ainda seja um desafio.

Neste trabalho, igualmente, esperamos parcerias para confirmar nosso resultados analíticos com simulações micro magnéticas, além de eventualmente estender nossas análises de controlabilidade com a Teoria do Controle Não-linear [22-24].

**Agradecimentos**